\documentclass[journal,twocolumn,10pt,letterpaper,final]{IEEEtran}

\usepackage{cite}
\usepackage{hyperref}
\ifCLASSINFOpdf
  \usepackage[pdftex]{graphicx}
  \graphicspath{{../pdf/}{../jpeg/}}
  \DeclareGraphicsExtensions{.pdf,.jpeg,.png,.eps}
\else
   \usepackage[dvips]{graphicx}
   \graphicspath{{../figures/}}
   \DeclareGraphicsExtensions{.eps,jpeg}
\fi

\usepackage[cmex10]{amsmath}
\usepackage{amssymb,amsthm}
\usepackage{dsfont}

\newtheorem{remark}{Remark}
\newtheorem{lemma}{Lemma}
\newtheorem{theorem}{Theorem}
\newtheorem{corollary}{Corollary}
\newtheorem{definition}{Definition}
\newtheorem{proposition}{Proposition}

\usepackage{array}
\usepackage{mdwmath}
\usepackage{mdwtab}
\usepackage{enumerate}
\usepackage{eqparbox}
\usepackage{mathtools}
\let\underbrace\LaTeXunderbrace

\usepackage[caption=false,font=footnotesize,labelfont=sf,textfont=sf]{subfig}

\usepackage{cases}

\renewcommand{\th}{^\text{th}}
\usepackage{fixltx2e}

\allowdisplaybreaks

\begin{document}
%
\title{Effect of Spatial Interference Correlation on the Performance of Maximum Ratio Combining}

\author{Ralph Tanbourgi\IEEEauthorrefmark{1},~\IEEEmembership{Student Member,~IEEE,} Harpreet S. Dhillon\IEEEauthorrefmark{2},~\IEEEmembership{Member,~IEEE,}\\Jeffrey G. Andrews\IEEEauthorrefmark{3},~\IEEEmembership{Fellow,~IEEE,} and
Friedrich K. Jondral\IEEEauthorrefmark{1},~\IEEEmembership{Senior Member,~IEEE}
\thanks{\IEEEauthorrefmark{1}R.~Tanbourgi and F.~K.~Jondral are with the Communications Engineering Lab (CEL), Karlsruhe Institute of Technology (KIT), Germany. Email: \texttt{\{ralph.tanbourgi, friedrich.jondral\}@kit.edu}. This work was supported by the German Academic Exchange Service (DAAD) and the German Research Foundation (DFG) within the Priority Program 1397 ``COIN'' under grant No. JO258/21-1.}
\thanks{\authorrefmark{2}H.~S.~Dhillon is with the Communication Sciences Institute (CSI), Department of Electrical Engineering, University of Southern California, Los Angeles, CA. Email: \texttt{hdhillon@usc.edu}.}
\thanks{\IEEEauthorrefmark{3}J.~G.~Andrews are with the Wireless and Networking Communications Group (WNCG), The University of Texas at Austin, TX, USA. Email: \texttt{dhillon@utexas.edu, jandrews@ece.utexas.edu}. This work was supported by NSF grant
CIF-1016649.}
}

\maketitle

\begin{abstract}
While the performance of maximum ratio combining (MRC) is well understood for a single isolated link, the same is not true in the presence of interference, which is typically correlated across antennas due to the common locations of interferers. For tractability, prior work focuses on the two extreme cases where the interference power across antennas is either assumed to be fully correlated or fully uncorrelated. In this paper, we address this shortcoming and characterize the performance of MRC in the presence of spatially-correlated interference across antennas. Modeling the interference field as a Poisson point process, we derive the exact distribution of the signal-to-interference ratio ($\mathtt{SIR}$) for the case of two receive antennas, and upper and lower bounds for the general case. Using these results, we study the diversity behavior of MRC and characterize the critical density of simultaneous transmissions for a given outage constraint. The exact $\mathtt{SIR}$ distribution is also useful in benchmarking simpler correlation models. We show that the full-correlation assumption is considerably pessimistic (up to $30\%$ higher outage probability for typical values) and the no-correlation assumption is significantly optimistic compared to the true performance. 
\end{abstract}


\IEEEpeerreviewmaketitle

\section{Introduction}\label{sec:introduction}

By exploiting the diversity provided by fading channels, multi-antenna receivers can enhance the communication performance. In the absence of multi-user interference or when interference is treated as white noise, it has been shown that MRC is optimal\cite{winters84,proakis95,aalo00,brennan03}. In MRC, the signals received at various branches or antennas are first weighted according to the signal-to-noise ratios experienced on those branches and then coherently combined to maximize the post-combiner signal-to-noise ratio. As with all the diversity-combining techniques, correlation among the signals received on different branches reduces the achievable diversity gains\cite{annavajjala04}, typically measured using the outage probability notion. For MRC in particular, fading correlation and average received-power imbalance across the branches, both of which are often encountered in practice, may reduce the resulting performance significantly when compared to the ideal case\cite{annavajjala04,aalo95}. Despite its sensitivity to such non-idealities, MRC is prevalent in most of today's wireless consumer products, such as wireless routers and laptops, that employ antenna-diversity.

\subsection{Related Work and Motivation}
In addition to the fading correlation, interference across diversity branches at a multi-antenna receiver is also {\em spatially correlated} due to the common locations of the interferers. Characterizing this type of correlation is challenging as it depends on many factors including the number and geometry of the surrounding interferers as well as their instantaneous channels towards a given receiver. Even worse, the network geometry, and hence the interference often appears random to the user due to mobility or irregular node deployment\cite{andrews10}, making a precise characterization of the resulting performance under spatial interference correlation cumbersome.

In this context, the authors of\cite{ganti09_1,haenggi12_1,gong12,Haenggi14twc,schilcher12} started using tools from stochastic geometry to obtain a more profound understanding of the interference correlation in a wireless network. These tools were identified as the key enablers for modeling the spatial and temporal interference correlation, and for analyzing their influence on various communications strategies. In principle, the interference is assumed to originate from a stochastic point process that models the interferer locations; thereby naturally capturing the origins of spatial correlation of interference. This approach led to an exact performance characterization of the simple retransmission mechanism\cite{ganti09_1} and of selection combining\cite{haenggi12_1} under interference correlation. An extension was presented in\cite{Haenggi14twc}, where it was shown that neglecting the interference correlation may lead to a delusive performance characterization. Similar tools were used in\cite{altieri11,tanbourgi_13_1,crismani13} to study the benefits of cooperative relaying in a multi-user scenario. These works clearly demonstrate that diversity exploiting techniques suffer a \emph{diversity loss} when interference correlation is properly accounted for. More sophisticated receive-diversity schemes that do not treat interference as pure noise were analyzed in\cite{govindasamy11} for linear minimum mean square error combining, and in\cite{ali10} for zero-forcing and optimal combining. The throughput scaling of decentralized networks with multi-antenna receivers was analyzed in\cite{jindal11}. 

Despite this progress, the performance characterization of MRC in the presence of spatial interference correlation is largely open and is the main focus of this paper. In\cite{hunter08,crismani13}, MRC was studied by assuming the \emph{same} interference level at all the receive branches which neglects the diversity in the fading gains of the interfering links. The effect of unequal interference levels on the outage probability for MRC was analyzed in\cite{aalo00,song05} for deterministic interference levels and without a specific correlation model. Instead of assuming the same (random) interference level across all receive antennas, the correlation may alternatively be completely \emph{neglected} as frequently done in the literature\cite[Chap.~3]{ganti09}; see\cite{sheng10} for an example with MRC.

Note that even though MRC is information-theoretically sub-optimal in the presence of interference\cite{cui99}, it is still of practical relevance since mass-market multi-antenna systems usually must treat interference as pure noise\cite{aalo00}, under which MRC achieves optimal performance. 

\subsection{Contributions and Outcomes}

In this paper, we characterize the distribution of $\mathtt{SIR}$ for MRC in the presence of spatially-correlated interference under realistic channel assumptions that include both long-term path loss effects and small scale fading effects, modeled as Rayleigh. The main contributions are summarized below.

{\em  Outage probability and the distribution of $\mathtt{SIR}$.}  
As the main result, we derive a closed-form expression for the cumulative distribution function (CDF) of the $\mathtt{SIR}$, equivalently outage probability, for the dual-antenna MRC case in Section~\ref{sec:main_result}. The result accounts for all relevant system parameters including transmitter density, path loss exponent and communication distance. For the important case of a path loss exponent of $4$, we obtain a simplified expression that requires only a single numerical integration. We stress that the dual-antenna case is of significant importance in current wireless systems, where most of the wireless devices, such as handhelds, laptops or wireless routers, are often equipped with at most two antennas due to complexity constraints and space limitations. In Section~\ref{sec:bounds}, we generalize our analysis to an arbitrary number of receive antennas by deriving lower and upper bounds on the $\mathtt{SIR}$ distribution. Although the construction of these bounds is rather simple, they allow a performance characterization of MRC. The usefulness of these bounds, quantified by the gap between the upper and lower bounds, decreases for very large number of antennas and small path loss exponents.

{\em Comparison with simpler correlation models.} The exact $\mathtt{SIR}$ distribution under spatial correlation can also be used to benchmark the performance of simpler correlation models used in the literature. We demonstrate that the full-correlation assumption for interference across receive branches yields a considerably pessimistic (up to roughly $30\%$ for typical values) estimate of the CDF of $\mathtt{SIR}$. This is because with the full-correlation assumption, the diversity among the fading gains on the different interfering links is effectively removed which, consequently, lowers the overall achievable diversity. In contrast, the no-correlation assumption overestimates the overall achievable diversity by neglecting the fact that the interference impinging at the different antennas originate from the same set of transmitters. As a result, the no-correlation assumption leads to a significantly optimistic characterization of the true performance.

{\em Applications of the developed theory.} In Section~\ref{sec:outage_scaling}, we characterize the diversity behavior of MRC using the notion of spatial-contention diversity order, which was introduced in\cite{tanbourgi_13_1}. While for a single isolated link, the diversity order gain (measured by the outage probability slope) of MRC theoretically scales with the number of antennas, this is not true for the multi-user case. This pitfall is due to the spatial interference correlation, which virtually disperses possible diversity order gains in the asymptotic regime. We then determine the network-wide critical density of simultaneous transmissions given a target outage probability in Section~\ref{sec:critical_density}. The exact critical density is obtained for the dual-antenna case using the main result, while the developed bounds are used to characterize the critical density for larger number of antennas. In order to complement the insights obtained using these bounds, we numerically estimate the true critical density and its scaling as a function of the number of receive antennas. A first-order approximation indicates a square-root dependence on the number of antennas.

\section{System Model}\label{sec:notation}

\begin{figure}[t]
	\centering
	\includegraphics[width=0.47\textwidth]{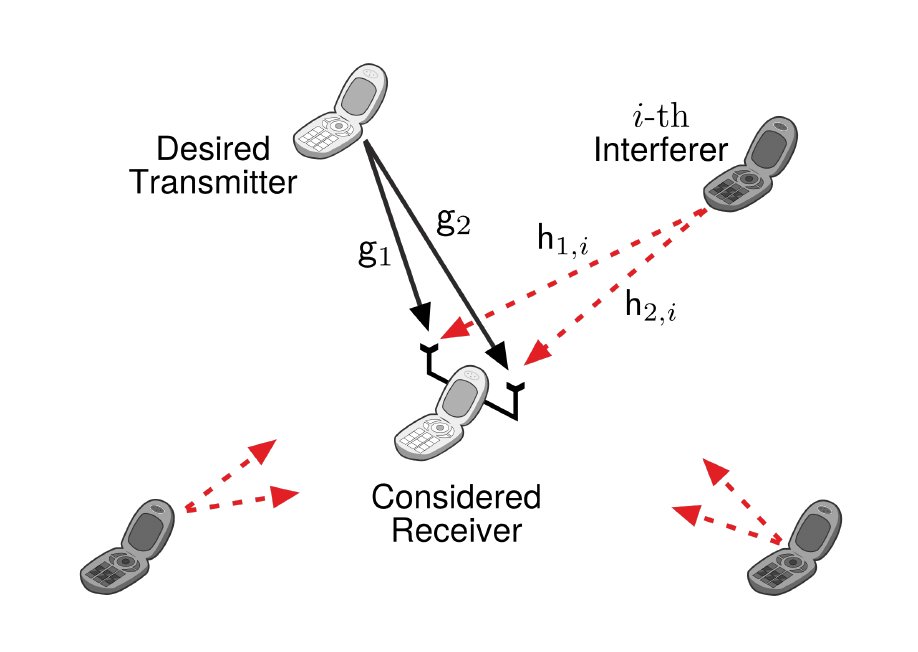}
	\caption{An illustration of the system model for $N=2$. The considered dual-antenna MRC receiver is located at the origin. The desired single-antenna transmitter is located $d$ meters away. Single-antenna interferers create interference to the considered dual-antenna MRC receiver.}
\label{fig:illustration}
\end{figure}

We consider an $N$-antenna receiver with a desired transmitter at distance $d$. The receiver experiences interference caused by other transmitters, whose locations $\{\mathsf{x}_{i}\}_{i=0}^{\infty}$ are modeled by a stationary Poisson point process (PPP) $\Phi\triangleq\{\mathsf{x}_{i}\}_{i=0}^{\infty}\subset\mathbb{R}^2$ of density $\lambda$. The density $\lambda$ influences the experienced interference and typical values for $\lambda$ range from $10^{-6}$, e.g., macro-cell deployment, to $10^{-2}$, e.g., dense WiFi deployment. The PPP assumption is widely-accepted\cite{andrews11,blas12} and provides a tractable way of dealing with spatial interference correlation. More complex interference geometries, e.g., carrier-sensing transmitter-inhibition, can be incorporated into the existing model using techniques from \cite{baccelli09b,hunter10,tanbourgi12}.\footnote{For other (non-Poisson) models and different fading, the form of the correlation might differ. Nevertheless, we expect the key insights in this work to be general and leave further extensions for possible future work.}

Because of the stationarity of $\Phi$\cite{stoyan95}, the interference experienced at a certain location is statistically the same at any other location. Thus, we can place the receiver under consideration in the origin $o\in\mathbb{R}^2$. The path loss between a point $x\in\mathbb{R}^2$ and the considered receiver is given by $\|x\|^{-\alpha}$, where $\alpha>2$ is the path loss exponent. We denote by $\mathsf{g}_{n}$ the channel fading power gain between the desired transmitter and the $n\th$ antenna of the considered receiver. Accordingly, we denote by $\mathsf{h}_{n,i}$ the channel fading power gain between the $i\th$ interferer and the $n\th$ antenna of the considered receiver. We further define $\mathbf{h}_{n}\triangleq\{\mathsf{h}_{n,i}\}_{i=0}^{\infty}$, $n\in[1,\ldots,N]$. We assume all fading gains to be independent and identically distributed (i.i.d.) with unit-mean exponential distribution, which models (narrow-band) Rayleigh fading. Possible extensions toward general fading distributions can be incorporated in the model, e.g., using ideas from\cite{keeler13,dhillon13}. We neglect noise and assume that transmissions are slotted and with fixed transmit power. The effect of (thermal) noise, variable transmit power and more complex medium access techniques is not treated in this work for better exposition of the main result. Their modeling as well as other extensions are left for possible future work. Figure~\ref{fig:illustration} illustrates the scenario for the example of a dual-antenna system.

We assume that the considered receiver can perfectly estimate the interference power level as well as the channel to the desired transmitter in every branch. By\cite{brennan03}, the combining weight for MRC in the $n\th$ branch is obtained by correcting the phase-mismatch of the received signal and scaling it by $\sqrt{\mathsf{g}_{n}}/\sum_{\mathsf{x}_{i}\in\Phi}\mathsf{h}_{n,i}\|\mathsf{x}_{i}\|^{-\alpha}$. Note that since this procedure is done for every branch independently of the others, possible common structure of the interference signals is not exploited, or equivalently, interference is treated as white noise in MRC. We can hence apply the same arguments as in the single-user case\cite{tse05}, yielding the post-combiner $\mathtt{SIR}$
\begin{IEEEeqnarray}{rCl}
	\mathtt{SIR}\triangleq\frac{\mathsf{g}_{1}d^{-\alpha}}{\sum\limits_{\mathsf{x}_{i}\in\Phi}\mathsf{h}_{1,i}\|\mathsf{x}_{i}\|^{-\alpha}}+\ldots+\frac{\mathsf{g}_{N}d^{-\alpha}}{\sum\limits_{\mathsf{x}_{i}\in\Phi}\mathsf{h}_{N,i}\|\mathsf{x}_{i}\|^{-\alpha}},\label{eq:sir_general}\IEEEeqnarraynumspace
\end{IEEEeqnarray}
where $\mathsf{g}_{n}d^{-\alpha}$ and $\sum_{\mathsf{x}_{i}\in\Phi}\mathsf{h}_{n,i}\|\mathsf{x}_{i}\|^{-\alpha}$ are the instantaneous received signal power and interference power in the $n\th$ branch, respectively. Now, the $\mathtt{SIR}$ is a random variable due to fading on the desired channels $\{\mathsf{g}_{n}\}_{n=1}^{N}$ \emph{and} due to the interference power levels (hereafter, interference), which depend on  $\{\mathbf{h}_{n}\}_{n=1}^{N}$ and $\Phi$.
Note that although all fading gains are assumed i.i.d., the $\mathtt{SIR}$s on the different branches are correlated as the interference terms originate from the same set of interferers $\Phi$.

{\em Notation:} Sans-serif-style letters ($\mathsf{z}$) denote random variables while serif-style letters ($z$) represent their realizations or variables. The function $(z)^{+}$ equals $z$ for $z>0$ and zero otherwise. We denote by $\mathbb{P}\left(\cdot\right)$ and $\mathbb{E}\left[\cdot\right]$ the probability measure and the expectation operator, respectively.

\section{Characterization of the \texorpdfstring{$\mathtt{SIR}$}{SIR}}\label{sec:sir}

This section is devoted to the characterization of the CDF of \eqref{eq:sir_general}. Our first main technical result is the exact CDF of $\mathtt{SIR}$ for the practically relevant case of two receive antennas ($N=2$). As will be evident from the derivation, there are several non-trivial challenges in this case which renders the general case of $N>2$ even more challenging. Therefore, we handle the case of $N>2$ by using bounding techniques.

\subsection{Exact Distribution of the \texorpdfstring{$\mathtt{SIR}$}{SIR} for \texorpdfstring{$N=2$}{N}}\label{sec:main_result}
In practice, wireless devices are often subject to complexity constraints and space limitations, thereby preventing the use of many antennas; for instance consumer electronics such as mobile handhelds, laptops or wireless routers are often equipped with no more than two antennas. It is therefore important to understand the particular case of $N=2$, for which the $\mathtt{SIR}$ reduces to
\begin{IEEEeqnarray}{rCl}
	\mathtt{SIR}&=&\frac{\mathsf{g}_{1}d^{-\alpha}}{\sum\limits_{\mathsf{x}_{i}\in\Phi}\mathsf{h}_{1,i}\|\mathsf{x}_{i}\|^{-\alpha}}+ \frac{\mathsf{g}_{2}d^{-\alpha}}{\sum\limits_{\mathsf{x}_{i}\in\Phi}\mathsf{h}_{2,i}\|\mathsf{x}_{i}\|^{-\alpha}}.
		\IEEEeqnarraynumspace\label{eq:sir_ccdf}
\end{IEEEeqnarray}
The CDF of $\mathtt{SIR}$ is an important quantity as it allows a detailed characterization of the link performance. For a given (coding/modulation-specific) $\mathtt{SIR}$ threshold $T$, the CDF can been seen as the outage probability. Equivalently, the complementary cumulative distribution function (CCDF) can be seen as the success probability ($1-$outage probability), which is characterized in the following Theorem.
\begin{theorem}\label{thm:cdf_general_two_antennas}
	The CCDF of $\mathtt{SIR}$ in the described setting for the case $N=2$ is given by
	\begin{IEEEeqnarray}{rCl}
	\mathbb{P}(\mathtt{SIR}\geq T)&=& 2\pi\lambda\int_{0}^{\infty}C(z,T)\int_{0}^{\infty}\frac{r^{-\alpha+1}d^{\alpha}}{(1+zr^{-\alpha}d^{\alpha})^2}\IEEEnonumber\\
	&&\qquad\qquad\times\frac{1}{1+r^{-\alpha}d^{\alpha}(T-z)^{+}}\,\mathrm dr\,\mathrm dz,\IEEEeqnarraynumspace\label{eq:general_two_antennas}
	\end{IEEEeqnarray}
where $C(z,T)$ is defined as
	\begin{IEEEeqnarray}{rCl}
	C(z,T) &\triangleq& \exp\left\{-2\pi\lambda\int_{0}^{\infty}r\,\left(1-\frac{1}{1+zr^{-\alpha}d^{\alpha}}\right.\right.\IEEEnonumber\\
	&&\qquad\qquad\qquad\left.\left.\times\frac{1}{1+r^{-\alpha}d^{\alpha}(T-z)^{+}}\right)\mathrm dr\right\}.\IEEEeqnarraynumspace\label{eq:general_two_antennas_sub}
	\end{IEEEeqnarray}
\end{theorem}

\begin{IEEEproof}
See Appendix~\ref{ap:general_two_antennas}.
\end{IEEEproof}

The result in Theorem~\ref{thm:cdf_general_two_antennas} requires the computation of three improper integrals. They can be numerically evaluated without difficulty using standard numeric software. For the special case $\alpha=4$, \eqref{eq:general_two_antennas_sub} reduces to closed form and \eqref{eq:general_two_antennas} requires only a single numerical integration. The result is given in Corollary~\ref{col:pathloss_four}.
\begin{corollary}\label{col:pathloss_four}
	For $\alpha=4$, the result of Theorem~\ref{thm:cdf_general_two_antennas} reduces to
	\begin{IEEEeqnarray}{rCl}
		\mathbb{P}(\mathtt{SIR}\geq T)&=&\frac{\pi^2}{4}d^2\lambda\int_{0}^{\infty}C_{4}(z,T)\IEEEnonumber\\
		&&\hspace{-.2cm}\times\frac{z^{\frac{3}{2}}-3\sqrt{z}(T-z)^{+}+2\left((T-z)^{+}\right)^{\frac{3}{2}}}{\left(z-(T-z)^{+}\right)^2}\,\mathrm dz,\IEEEeqnarraynumspace\label{eq:cor_alpha4}
	\end{IEEEeqnarray}
where $C_{4}(z,T)$ is defined as
\begin{IEEEeqnarray}{rCl}
	C_{4}(z,T)\triangleq\exp\left(-\frac{\pi^2}{2}\lambda d^2\frac{z^{\frac{3}{2}}-\left((T-z)^{+}\right)^{\frac{3}{2}}}{z-(T-z)^{+}}\right).
\end{IEEEeqnarray}
\end{corollary}

Note that the case $\alpha=4$ is frequently found in outdoor wireless systems because of ground plane reflection effects in the wireless channel\cite[Chap.~2]{tse05}.

\begin{figure}[!t]
	\centering
	\includegraphics[width=0.48\textwidth]{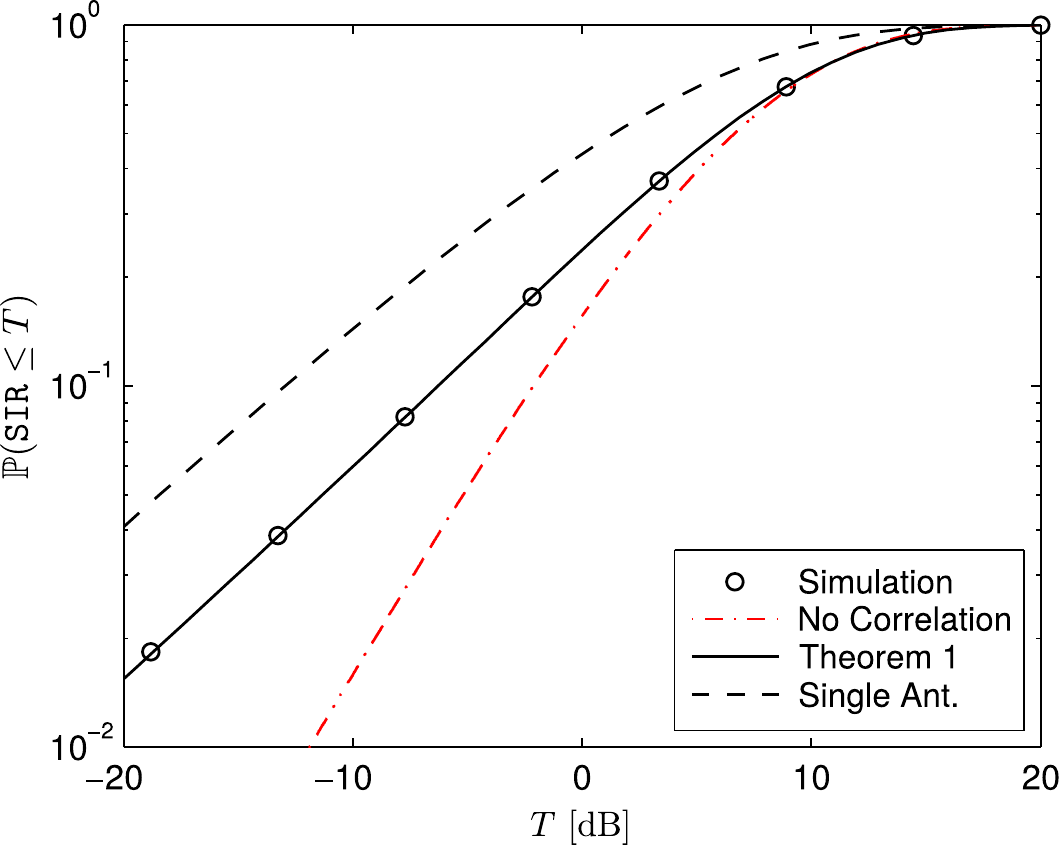}
	\caption{$\mathtt{SIR}$ CDF with parameters $\lambda=10^{-3}$, $\alpha=3.5$, $d=10$, $N=2$.}
	\label{fig:valid_log}
\end{figure}

Figure~\ref{fig:valid_log} shows the CDF of $\mathtt{SIR}$ for different values of $T$. First of all, it can be seen that the theoretical result from Theorem~\ref{thm:cdf_general_two_antennas} matches the simulation results perfectly. The dotted-dashed line illustrates the expected performance if no spatial correlation was assumed. This scenario was obtained by creating two interference realizations independently of each other in the simulation. It is clear that the no-correlation assumption is by far too optimistic and does not recover the true order of decay of $\mathbb{P}(\mathtt{SIR}\leq T)$ at small $T$.  

\subsection{Simple Bounds on the \texorpdfstring{$\mathtt{SIR}$}{SIR} for Arbitrary \texorpdfstring{$N$}{N}}\label{sec:bounds}

Although the case $N=2$ already covers a broad range of practical scenarios, it would be interesting to characterize the performance of MRC also for $N>2$. Since the exact characterization is clearly challenging, we proceed by deriving various useful bounds.

\subsubsection{Full-correlation assumption}

A commonly made assumption when analyzing diversity-combining is to assume that the interference realizations in the different branches are the same, i.e., the interference is \emph{fully-correlated} among the branches, see for instance\cite{hunter08,crismani13}. This is, however, not true in general since each interference signal might undergo a fading realization that is different for each receive antenna. The full-correlation assumption is formalized as follows.

\begin{definition}[Full-correlation (FC) assumption]
	Under the FC assumption, the interference terms $\sum_{\mathsf{x}_{i}\in\Phi}\mathsf{h}_{n,i}\|\mathsf{x}_{i}\|^{-\alpha}$ at the $N$ antennas are assumed to be equal, i.e., $\mathsf{h}_{m,i}\equiv\mathsf{h}_{n,i}$ for all $m,n\in[1,\ldots,N]$ and $i\in\mathbb{N}$. The corresponding $\mathtt{SIR}$ is denoted by $\mathtt{SIR}_{\text{\textnormal{FC}}}\triangleq (d^{-\alpha}\sum_{n=1}^{N}\mathsf{g}_n)/\sum_{\mathsf{x}_{i}\in\Phi}\mathsf{h}_{1,i}\|\mathsf{x}_{i}\|^{-\alpha}$.
\end{definition}

The reason for which the FC assumption is included in this work is two-fold: first, it would be interesting to study the gap to the exact result (which is now available for $N=2$). Second, it turns out that the FC assumption provides an upper bound on the exact CDF of the $\mathtt{SIR}$. Before proceeding, we note the following useful Lemma.\nopagebreak [4] 

\begin{lemma}\label{lem:gamma_interference} Let $\mathsf{U}$ be a random variable and denote by $ \mathcal{L}_{\mathsf{U}}(s)$ the Laplace transform of the probability density function (PDF) of $\mathsf{U}$ and by $\partial^k\mathcal{L}_{\mathsf{U}}(s)/\partial s^{k}$ its $k$-th derivative. Then,
	\begin{IEEEeqnarray}{rCl}
		\mathbb{P}\left(\frac{\mathsf{g}_{1}+\ldots+\mathsf{g}_{N}}{\mathsf{U}d^{\alpha}}\geq v\right)
			=\sum\limits_{k=0}^{N-1}(-1)^{k}\frac{s^{k}}{k!}\,\frac{\partial^{k} \mathcal{L}_{\mathsf{U}}(s)}{\partial s^{k}}\Big\lvert_{s=vd^{\alpha}}.\IEEEeqnarraynumspace\label{eq:gamma_interference}
	\end{IEEEeqnarray}
\end{lemma}
\begin{IEEEproof}
	We write
	\begin{IEEEeqnarray}{rCl}
	&&\hspace{-.6cm}\mathbb{P}\left(\mathsf{g}_{1}+\ldots+\mathsf{g}_{N}\geq vd^{\alpha}\mathsf{U}\right)\IEEEnonumber\\
	&\overset{\text{(a)}}{=}&\mathbb{E}\left[\frac{\Gamma(N,vd^{\alpha}\mathsf{U})}{(N-1)!}\right]\IEEEnonumber\\
			&\overset{\text{(b)}}{=}&\int_{0}^{\infty}e^{-vd^{\alpha}u}\sum\limits_{k=0}^{N-1}\frac{(vd^{\alpha})^{k}}{k!}u^{k}\,\mathrm d\mathbb{P}\left(\mathsf{U}\leq u\right)\IEEEnonumber\\
			&\overset{\text{(c)}}{=}&\sum\limits_{k=0}^{N-1}(-1)^{k}\frac{(vd^{\alpha})^{k}}{k!}\,\underbrace{(-1)^{k}\int_{0}^{\infty}u^{k}e^{-vd^{\alpha}u}\,\mathrm d\mathbb{P}\left(\mathsf{U}\leq u\right)}_{\frac{\partial^{k} \mathcal{L}_{\mathsf{U}}(s)}{\partial s^{k}}\lvert_{s=vd^{\alpha}}},\IEEEeqnarraynumspace
	\end{IEEEeqnarray}
	where (a) follows from conditioning on $\mathsf{J}$ and noting that $\mathsf{g}_{1}+\ldots+\mathsf{g}_{N}$ is $\Gamma$-distributed with shape $N$ and unit scale. (b) follows from the relation $\Gamma(b,x)=(b-1)!e^{-x}\sum_{k=0}^{b-1}\frac{x^k}{k!}$ for positive integer $b$, and (c) is obtained by interchanging integration and summation which is allowed by the dominated convergence theorem. Alternatively, one can set  $n=1$ and $a_{nk}=1/k!$ in\cite[Theorem~1]{hunter08} to obtain the same result.
\end{IEEEproof}

The $k$-th derivative in \eqref{eq:gamma_interference} can be efficiently computed using Fa\`{a} di Bruno's rule\cite{bruno1857} together with Bell polynomials\cite{johnson07}. 

\begin{proposition}\label{prop:fc}
	The CCDF of $\mathtt{SIR}_{\text{\textnormal{FC}}}$ is given by
	\begin{IEEEeqnarray}{rCl}
		\mathbb{P}\left(\mathtt{SIR}_{\text{\textnormal{FC}}}\geq T\right)=\sum\limits_{k=0}^{N-1}(-1)^{k}\frac{s^{k}}{k!}\frac{\partial^{k}}{\partial s^{k}}e^{-c s^{\frac{2}{\alpha}}}\Big\lvert_{s=Td^{\alpha}},\IEEEeqnarraynumspace\label{eq:general_fc}
	\end{IEEEeqnarray}
	where $c=\tfrac{2}{\alpha}\pi^2\lambda {\rm csc}(2\pi/\alpha)$. For the special case $N=2$, \eqref{eq:general_fc} can be simplified to
	\begin{IEEEeqnarray}{rCl}
		\mathbb{P}\left(\mathtt{SIR}_{\text{\textnormal{FC}}}\geq T\right)=e^{-c d^{2}T^{\frac{2}{\alpha}}}\left(1-\tfrac{2}{\alpha}cd^{2}T^{\frac{2}{\alpha}}\right).\IEEEeqnarraynumspace\label{eq:two_antenna_fc}
	\end{IEEEeqnarray}
\end{proposition}

\begin{figure}[t]
	\centering
	\includegraphics[width=0.48\textwidth]{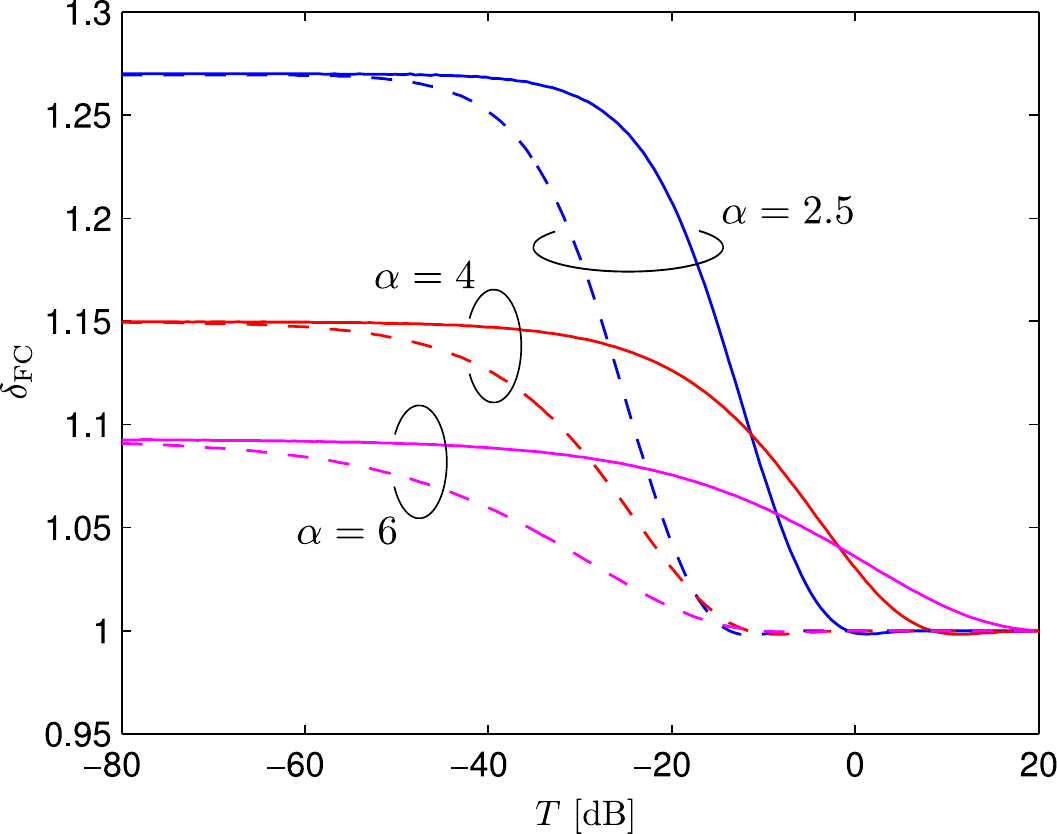}
	\caption{Deviation $\delta_{\text{FC}}$ vs. $T$ for different $\alpha,\,\lambda$. Parameters are: $N=2$, $d=15$, $\lambda=10^{-2}$ (dashed line), $\lambda=10^{-3}$ (solid line).}
	\label{fig:gap_theory_fc}
\end{figure}

Figure~\ref{fig:gap_theory_fc} illustrates the CDF deviation $\delta_{\text{FC}}\triangleq\mathbb{P}(\mathtt{SIR}_{\text{FC}}\leq T)/\mathbb{P}(\mathtt{SIR}\leq T)$ vs. $T$ for different $\alpha,\,\lambda$. The results were obtained by computing the CDFs using \eqref{eq:general_two_antennas}, \eqref{eq:general_two_antennas_sub} and \eqref{eq:two_antenna_fc}. It can be seen that the deviation becomes large for small $T$. Interestingly, for asymptotically small $T$, this gap solely depends on the path loss exponent with values roughly between $8\%$ to $27\%$ for typical system parameters. The points at which the lines hit the value one (negligible deviation) correspond to the $T$-values at which $\mathbb{P}(\mathtt{SIR}\leq T)$ is roughly $0.9$. For values beyond $0.9$ (non-practical regime) the FC assumption becomes a lower bound on the exact CDF of $\mathtt{SIR}$.

\begin{remark}[Upper bound on the $\mathtt{SIR}$ CDF]
It is intuitive that the FC assumption yields an upper bound on the $\mathtt{SIR}$ CDF due to the fact that the additional correlation in the fading gains of the interfering links decreases the channel variability, and hence the obtainable diversity\cite{annavajjala04}. From this observation, we thus conjecture that the FC assumption provides an upper bound on the CDF of the $\mathtt{SIR}$ also for a larger number of antennas $N$. Simulation results support this conjecture.
\end{remark}

\subsubsection{Max/min-fading based bounds}

Simple bounds can be constructed by modifying the statistics of the fading gains $\{\mathbf{h}_{n}\}_{n=1}^{N}$ in the following way.
\begin{definition}[$\max$/$\min$-fading]
	For $\max$-fading, the gains $\mathsf{h}_{n,i}$ are set according to the rule $\mathsf{h}_{n,i}\equiv\mathsf{h}_{\max,i}\equiv\max_{k}\{\mathsf{h}_{k,i}\}$ for all $n\in[1,\ldots,N]$ and $i\in\mathbb{N}$. Similarly, the gains for $\min$-fading are set according to $\mathsf{h}_{n,i}\equiv\mathsf{h}_{\min,i}\equiv\min_{k}\{\mathsf{h}_{k,i}\}$ for all $n\in[1,\ldots,N]$ and $i\in\mathbb{N}$. The respective $\mathtt{SIR}$s are denoted by $\mathtt{SIR}_{\max}\triangleq (d^{-\alpha}\sum_{n=1}^{N}\mathsf{g}_n)/\sum_{\mathsf{x}_{i}\in\Phi}\mathsf{h}_{\max,i}\|\mathsf{x}_{i}\|^{-\alpha}$ and $\mathtt{SIR}_{\min}\triangleq (d^{-\alpha}\sum_{n=1}^{N}\mathsf{g}_n)/\sum_{\mathsf{x}_{i}\in\Phi}\mathsf{h}_{\min,i}\|\mathsf{x}_{i}\|^{-\alpha}$. 
\end{definition}

\begin{proposition}\label{prop:minmax_bounds}
	In the described setting,
	\begin{IEEEeqnarray}{rCl}
	&&\mathbb{P}\left(\mathtt{SIR}_{\min}\geq T\right)\IEEEnonumber\\
	&&\hspace{.1cm}=\sum\limits_{k=0}^{N-1}(-1)^{k}\frac{s^{k}}{k!}\frac{\partial^{k}}{\partial s^{k}}\exp\left\{-\tfrac{2}{\alpha}\pi^2\lambda s^{\frac{2}{\alpha}}\mathrm{csc}\left(\tfrac{2\pi}{\alpha}\right)\right\}\Big\lvert_{s=\frac{T}{N}d^{\alpha}}\IEEEeqnarraynumspace
	\end{IEEEeqnarray}
	and
	\begin{IEEEeqnarray}{rCl}
	&&\mathbb{P}\left(\mathtt{SIR}_{\max}\geq T\right)=\sum\limits_{k=0}^{N-1}(-1)^{k}\frac{s^{k}}{k!}\IEEEnonumber\\
	&&\quad\quad\times\frac{\partial^{k}}{\partial s^{k}}\exp\left\{-\lambda\pi s^{\frac{2}{\alpha}}\Gamma(1-\tfrac{2}{\alpha})\mathbb{E}\left[\mathsf{h}_{\max}^{\frac{2}{\alpha}}\right]\right\}\Big\lvert_{s=Td^{\alpha}},\IEEEeqnarraynumspace
	\end{IEEEeqnarray}
	where $\mathsf{h}_{\max}$ has distribution $\mathbb{P}\left(\mathsf{h}_{\max}\leq h\right)=(1-\exp(-h))^{N}$. Furthermore,
	\begin{IEEEeqnarray}{rCl}
		\mathbb{P}\left(\mathtt{SIR}_{\min}\geq T\right)\geq\mathbb{P}\left(\mathtt{SIR}\geq T\right)\geq\mathbb{P}\left(\mathtt{SIR}_{\max}\geq T\right).\IEEEeqnarraynumspace\label{eq:minmax_bounds}
	\end{IEEEeqnarray}
\end{proposition}
\begin{IEEEproof}
	See Appendix~\ref{ap:prop_minmax_bounds}.
\end{IEEEproof}

The result of Proposition~\ref{prop:minmax_bounds} can be further simplified for cases of special interest.
\begin{corollary}
	For $N=2$ the result in \eqref{eq:minmax_bounds} yields
	\begin{IEEEeqnarray}{rCl}
		\mathbb{P}\left(\mathtt{SIR}\geq T\right)\overset{c=c_{1}}{\underset{c=c_{2}}\gtreqless}
			\frac{e^{-cd^{2}T^{\frac{2}{\alpha}}}}{\alpha}\left(\alpha+2cd^{2}T^{\frac{2}{\alpha}}\right),\IEEEeqnarraynumspace
	\end{IEEEeqnarray}
	where $c_1=\tfrac{2^{1-\frac{2}{\alpha}}}{\alpha}\pi^2\lambda \mathrm{csc}\left(\tfrac{2\pi}{\alpha}\right)$ and $c_2=\tfrac{4-2^{1-\frac{2}{\alpha}}}{\alpha}\lambda\pi^2 \mathrm{csc}\left(\tfrac{2\pi}{\alpha}\right)$.
\end{corollary}
\begin{corollary}
	For $N=4$ the result in \eqref{eq:minmax_bounds} yields
		\begin{IEEEeqnarray}{rCl}
		&&\mathbb{P}\left(\mathtt{SIR}\geq T\right)\IEEEnonumber\\
		&&\quad\quad\overset{c=c_{1}}{\underset{c=c_{2}}\gtreqless}
			\frac{e^{-cT^{\frac{2}{\alpha}}}}{3\alpha^3}\left(3\alpha^3+11\alpha^2cT^{\frac{2}{\alpha}}+12\alpha cT^{\frac{2}{\alpha}}(cT^{\frac{2}{\alpha}}-1)\right.\IEEEnonumber\\
			&&\hspace{3.3cm}\left.+4 c T^{\frac{2}{\alpha}}(1-3cT^{\frac{2}{\alpha}}+c^2T^{\frac{4}{\alpha}})\right),\IEEEeqnarraynumspace
	\end{IEEEeqnarray}
	where $c_1=\tfrac{2^{1-\frac{4}{\alpha}}}{\alpha}\pi^2\lambda d^{2}\mathrm{csc}\left(\tfrac{2\pi}{\alpha}\right)$ and $c_2=(8-3\times 2^{2-\frac{2}{\alpha}}-2^{1-\frac{4}{\alpha}}+8\times3^{-\frac{2}{\alpha}})\frac{\pi^2}{\alpha}\lambda d^{2}\mathrm{csc}\left(\tfrac{2\pi}{\alpha}\right)$. 
\end{corollary}

For instance, when $\alpha=4$, we have $c_{1}=.25\pi^2\lambda d^2$ and $c_{2}=.78\pi^2\lambda d^2$ for the case $N=4$. When the product $\pi\lambda d^2 T^{\frac{2}{\alpha}}\Gamma(1-\tfrac{2}{\alpha})$ is small, \eqref{eq:minmax_bounds} can be further simplified using a Taylor series expansion of the $\exp$ term as shown next.
\begin{corollary}\label{cor:simple_minmax}
	Let $c'=\pi\lambda d^2 T^{\frac{2}{\alpha}}\Gamma(1-\tfrac{2}{\alpha})$. As $c'\to0$, we have
	\begin{IEEEeqnarray}{rCl}
		 N^{-\frac{2}{\alpha}}\Gamma(1+\tfrac{2}{\alpha}) D(\alpha,N)&\leq&\lim_{c'\to0}\frac{1}{c'}\,\mathbb{P}\left(\mathtt{SIR}\leq T\right)\IEEEnonumber\\
		 &\leq&\mathbb{E}\left[\mathsf{h}_{\max}^{\frac{2}{\alpha}}\right] D(\alpha,N),\IEEEeqnarraynumspace\label{eq:simplified_minmax}
	\end{IEEEeqnarray}
	where $D(\alpha,N)=\sum_{k=0}^{N-1}\frac{(-1)^{k}}{k!}(1+\tfrac{2}{\alpha}-k)_{k}$ and $(a)_{k}$ being the Pochhammer symbol\cite{olver10}.
\end{corollary}

\begin{figure}[t]
	\centering
	\includegraphics[width=0.48\textwidth]{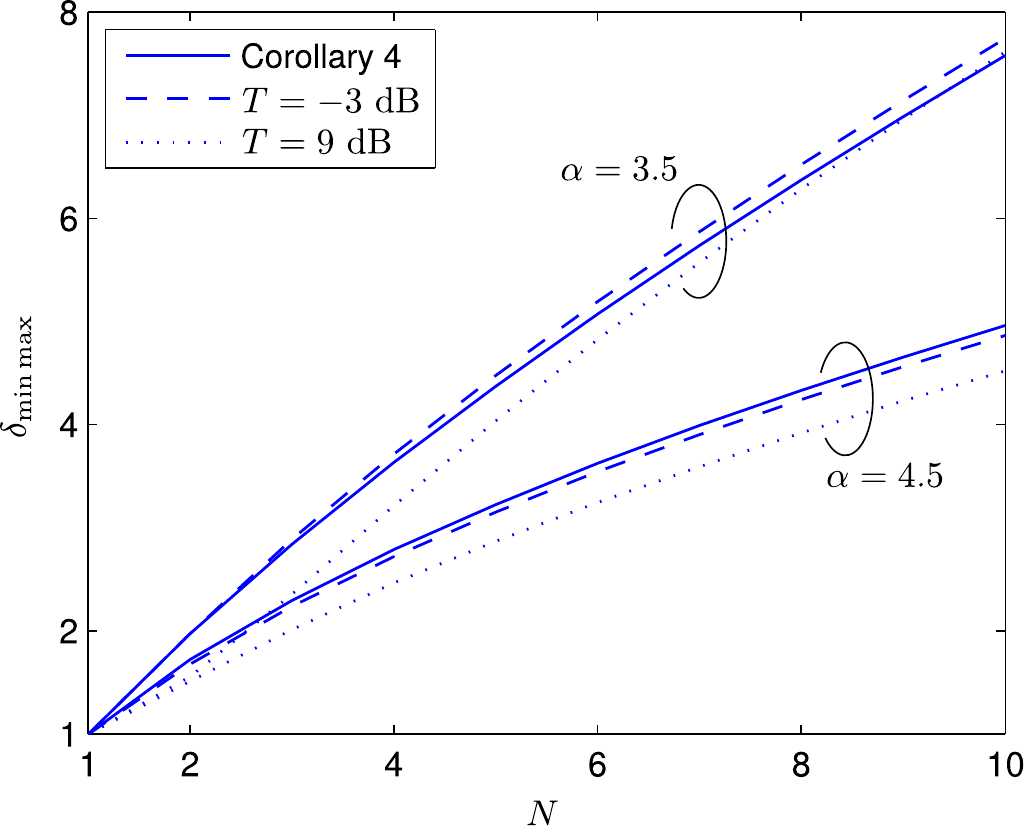}
	\caption{Gap $\delta_{\min\max}$ for different $\alpha,N$. Asymptotic $\delta_{\min\max}$ (solid) is obtained using Corollary~\ref{cor:simple_minmax}. Non-asymptotic $\delta_{\min\max}$ shown for $T=-3$~dB (dashed) and $T=9$~dB (dotted) with parameters $\lambda=10^{-3}$, $d=10$.}
	\label{fig:minmax_gap}
\end{figure}

\begin{figure*}[!t]
	\centerline{\subfloat[$\mathtt{SIR}$ CDF for various models]{\includegraphics[width=0.472\textwidth]{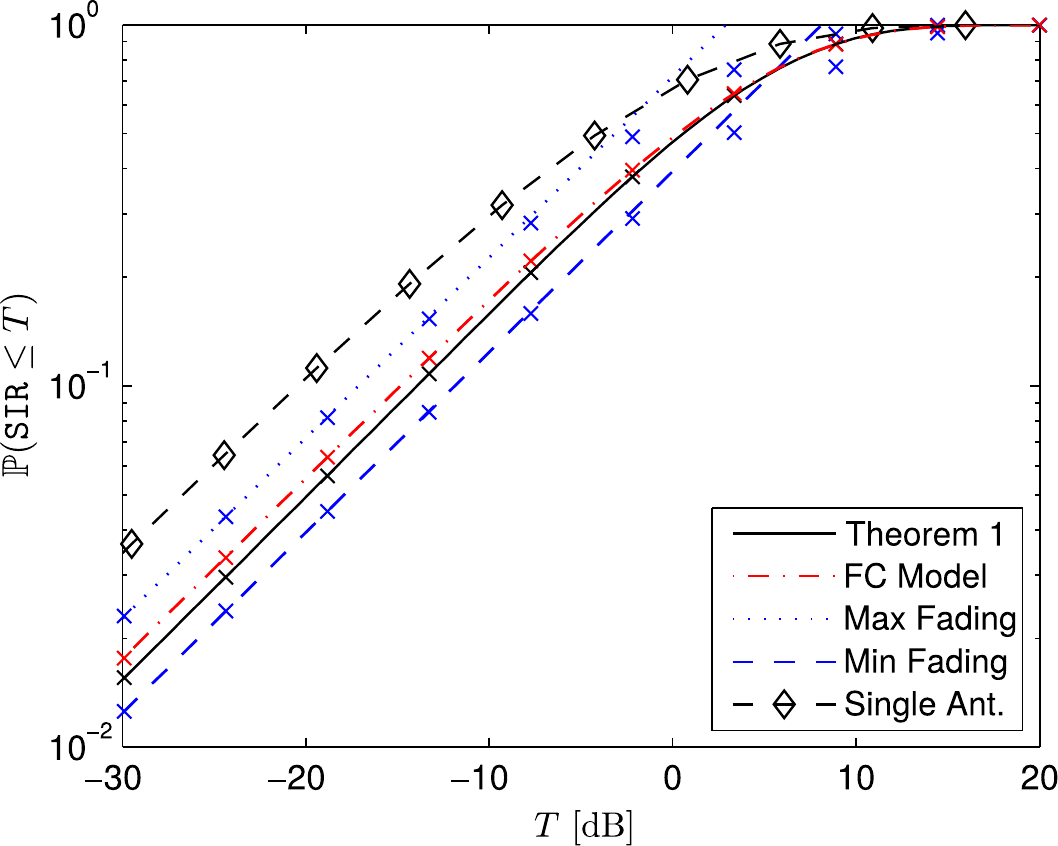}\label{fig:bnd_log}}
	\hfil
	\subfloat[Asymptotic $\mathtt{SIR}$ CDF]{\includegraphics[width=0.48\textwidth]{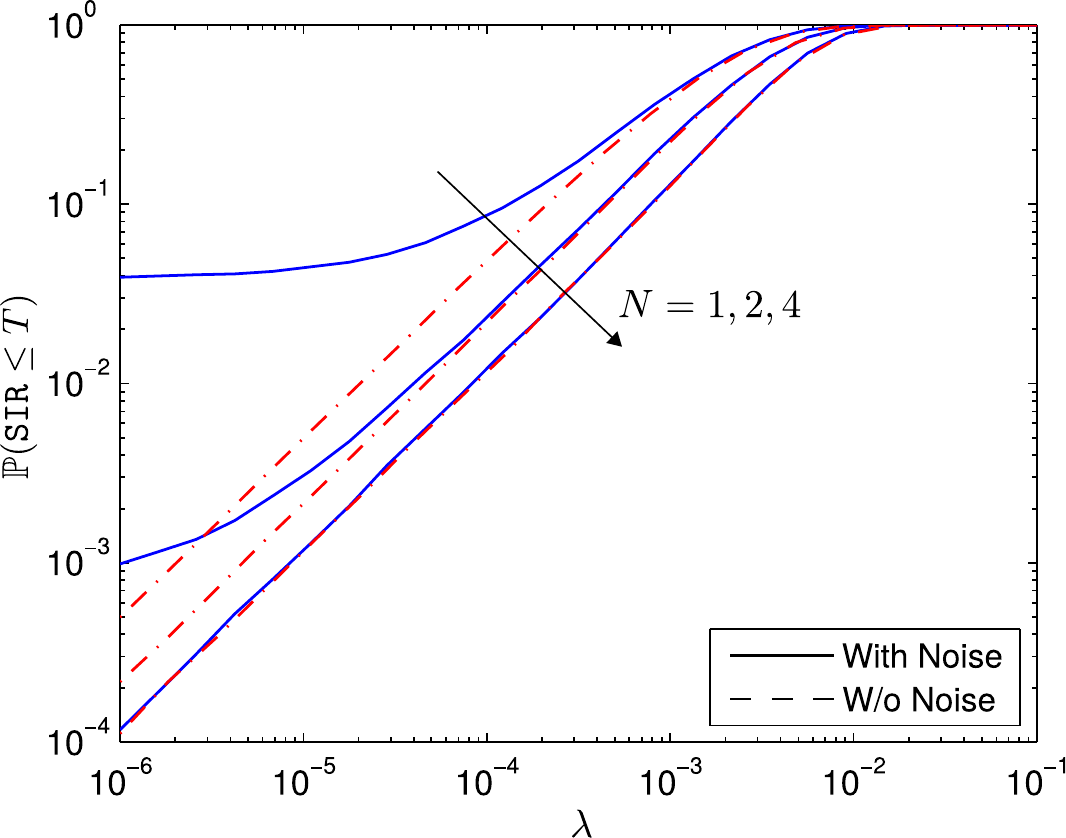}\label{fig:conjecture2}}}
	\caption{(a) $\mathbb{P}(\mathtt{SIR}\leq T)$ vs. $T$. Parameter: $\lambda=10^{-3}$, $\alpha=4$, $d=15$, $N=2$. ``x''-marks represent simulation results. (b) Simulated CDF of $\mathtt{SIR}$ vs. $\lambda$ for different $N$ with (solid) and without (dashed) additional noise. Parameters are $\alpha=4$, $d=10$, $T=1$. Average SNR is 14 dB (target outage probability of $4\%$ for $N=1$).}
\end{figure*}

The gap $\delta_{\min\max}\triangleq\mathbb{P}(\mathtt{SIR}_{\max}\leq T)/\mathbb{P}(\mathtt{SIR}_{\min}\leq T)$ is shown in Fig.~\ref{fig:minmax_gap}. The solid line (Corollary~\ref{cor:simple_minmax}) follows from dividing the upper bound by the lower bound in \eqref{eq:simplified_minmax}. It can be seen that $\delta_{\min\max}$ increases with the number of antennas $N$ and/or with decreasing $\alpha$. Fig.~\ref{fig:bnd_log} shows $\mathbb{P}(\mathtt{SIR}\leq T)$ vs. $T$ for the various expressions obtained in Section~\ref{sec:bounds} together with the exact result (Theorem~\ref{thm:cdf_general_two_antennas}, solid) and the single-antenna case (dashed+diamonds). The dotted-dashed line corresponds to the FC assumption (Proposition~\ref{prop:fc}), while the dashed and dotted lines correspond to the $\min$- and $\max$-fading bounds, respectively (Corollary~\ref{cor:simple_minmax}). The ``x''-marks represent the simulation results. The figure suggests that the FC assumption yields a tighter upper bound on the $\mathtt{SIR}$ CDF compared to the $\max$-fading bound.

\section{Applications}\label{sec:applications}

\subsection{Outage Probability Scaling with \texorpdfstring{$\lambda$}{Density}}\label{sec:outage_scaling}
The diversity order\cite{laneman04} is an asymptotic metric for quantifying the gains of diversity techniques in the interference-free high-reliability regime, i.e., at small outage probabilities. While in the single-user case this regime is typically achieved by scaling the transmit power, this is not true for the multi-user case; jointly increasing transmit power does not change the $\mathtt{SIR}$. In (decentralized) multi-user systems, efficient MAC protocols usually control the \emph{density} of concurrent transmissions to achieve a sufficiently high $\mathtt{SIR}$, e.g., Aloha (spatial reuse with a medium access probability) and carrier sense multiple access (spatial inhibition of simultaneously active transmitters). It is therefore interesting to analyze the achievable diversity order as a function of $\lambda$.

The spatial-contention diversity order (SC-DO) was introduced in\cite{tanbourgi_13_1} and is defined as
\begin{IEEEeqnarray}{c}
	\Delta\triangleq\lim_{\lambda\to 0} \frac{\log \mathbb{P}(\mathtt{SIR}\leq T)}{\log \lambda}
\end{IEEEeqnarray}
for $T\in(0,\infty)$. It characterizes the slope of the outage probability when letting the density of transmitters $\lambda$ tend to zero, and hence --  similar to the diversity order metric -- quantifies the diversity gain in the high-reliability regime. As an example, for $N=1$ the outage probability is $1-\exp(-\tfrac{2}{\alpha}\lambda\pi^2T^{\frac{2}{\alpha}}d^2\csc(2\pi/\alpha))$\cite{baccelli06,ganti09,weberandrews11}. The corresponding SC-DO is $\Delta=1$ as expected. It remains to be clarified whether the SC-DO is larger than one for $N>1$ with MRC in the described setting.

\begin{theorem}\label{thm:sc-do}
	The SC-DO in the described setting for the case $N=2$ is $\Delta=1$.
\end{theorem}
\begin{IEEEproof}
See Appendix~\ref{ap:sc-do}.
\end{IEEEproof}

This result is consistent with the findings obtained in\cite{tanbourgi_13_1}, where it was shown that there is no diversity order gain with respect to the density $\lambda$ as a result of the spatial interference correlation. It is important to note that the SC-DO is a useful diversity metric only in the interference-limited regime since, although noise is neglected in the model, letting $\lambda\to0$ would render performance noise-limited in practice. To elucidate this note more, Fig.~\ref{fig:conjecture2} shows the simulated outage probability vs. $\lambda$ with and without additional noise. It can be seen that, depending on the number of antennas $N$, the noise-free asymptotic slope is observed in the noise-added case only in limited intervals, e.g., when $\lambda$ is approximately between $10^{-5}$ and $10^{-3}$ for $N=2$. In these intervals, the SC-DO is a useful metric for characterizing the diversity gains. Note that although Theorem~\ref{thm:sc-do} treats only the case $N=2$, it is reasonable to conjecture that adding more antennas will not change the SC-DO $\Delta=1$. Fig.~\ref{fig:conjecture2} supports this conjecture.

\begin{remark}[Diversity order definition]
The diversity order for interference-limited networks can alternatively be defined in a more general way by considering the limit $\mathtt{SIR}\to\infty$, which can be achieved in different ways, including $\lambda\to0$, as shown in\cite{Haenggi14twc}. From the mapping theorem\cite[Theorem~2.1]{weberandrews11} it follows that $\mathtt{SIR}\propto\lambda^{-\frac{\alpha}{2}}$ which links the result of Theorem~\ref{thm:sc-do} to the diversity order results in\cite{Haenggi14twc}. 
\end{remark}

\subsection{Critical Density}\label{sec:critical_density}

From the results obtained in Section~\ref{sec:main_result} and Section~\ref{sec:bounds} it is apparent that adding more nodes increases the interference, and hence worsens the $\mathtt{SIR}$. In decentralized networks it is desirable to know the number of transmissions per unit area that can be supported over a target distance $d$ subject to a quality-of-service constraint. The target-distance assumption is known in the literature as the ``dipole model''\cite[Chap.~16.2]{baccelli09b} and is commonly used for characterizing the spatial throughput in wireless networks, cf.\cite{weber10}. Given a target outage probability $\epsilon\triangleq\mathbb{P}(\mathtt{SIR}<T)$, the \emph{critical density} $\lambda_{\epsilon}$ gives the maximum allowable density of simultaneous transmissions over distance $d$ with probability of failure $\epsilon$. It can be obtained by solving $\mathbb{P}(\mathtt{SIR}<T)$ for $\lambda$. In our case, numerical methods have to be used due to the nested structure of the CDF in \eqref{eq:general_two_antennas}.

\begin{figure}[!t]
	\centering
	\includegraphics[width=0.48\textwidth]{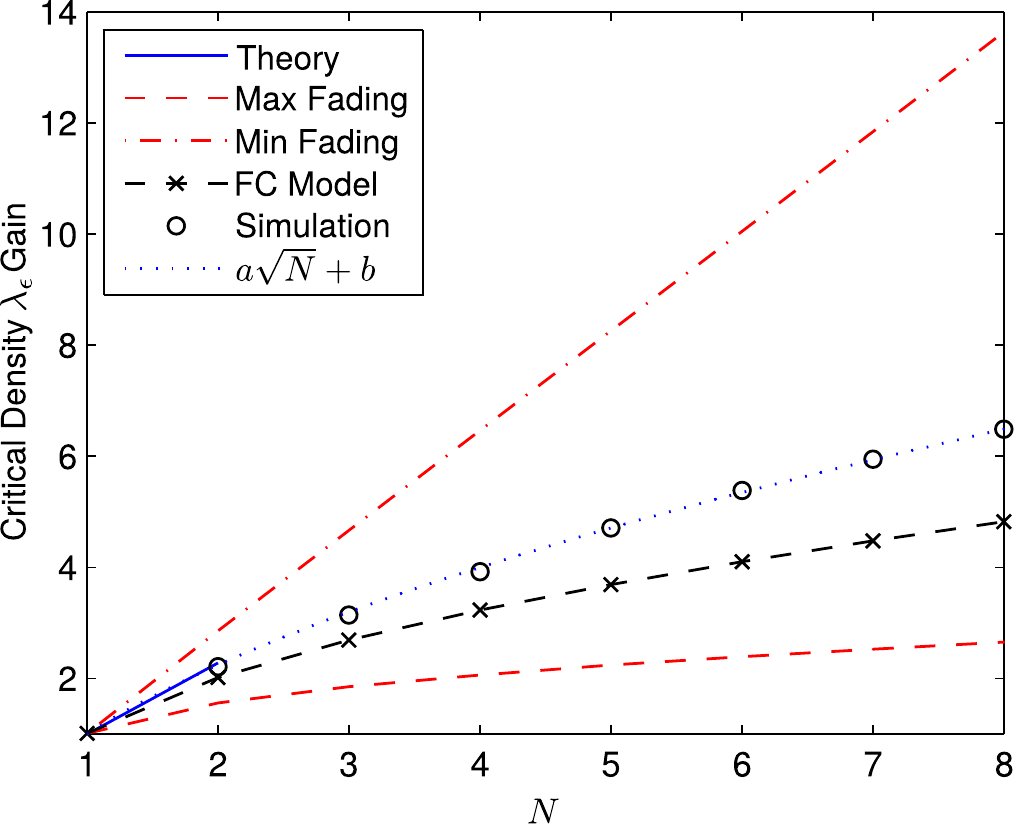}
	\caption{Critical density $\lambda_{\epsilon}$ gain over single-antenna systems vs. $N$. Parameters are $\epsilon=0.05$, $\alpha=4$, $d=15$, $T=1$. In this example, $a=3$ and $b=-2$ were chosen as the curve fitting coefficients.}
	\label{fig:critical_density}
\end{figure}

Figure~\ref{fig:critical_density} shows the critical density $\lambda_{\epsilon}$ gain over single-antenna systems for different $N$. The critical density $\lambda_{\epsilon}$ in the single-antenna case is given by $\lambda_{\epsilon}=\tfrac{-\alpha\log(1-\epsilon)}{2\pi^2 d^2\text{csc}(2\pi/\alpha)T^{2/\alpha}}$\cite{weber10}. For reference, we also included the corresponding curves for $\min$-/$\max$-fading and FC-correlation for the high-reliability regime. While these curves may help in bounding the critical density for larger $N$, Fig.~\ref{fig:critical_density} demonstrates the importance for studying the critical density under more realistic assumptions as none of the curves tend to provide the true scaling behavior.

\begin{remark}[Scaling of $\lambda_{\epsilon}$ with $N$]
Figure~\ref{fig:critical_density} reveals a sublinear growth of the critical density as the number of antennas increases. A first-order approximation indicates that the scaling is proportional to $\sqrt{N}$.
\end{remark}

\section{Conclusion}\label{sec:conclusion}
In contrast to the single-user scenario, the performance of MRC in a multi-user scenario is not well understood, primarily due to the presence of spatial correlation in the interference across diversity branches. In this work, we addressed this shortcoming and derived the exact CDF of the $\mathtt{SIR}$ for dual-antenna MRC in the presence of spatially-correlated interference. The result is given in form of easy-to-solve integrals, which can be further simplified in certain special cases of interest. This result covers a large range of practical applications and offers valuable insights: (i) when the spatial correlation of the interference is factored in, MRC does not change the outage probability slope over the interferer density in the high-reliability regime; (ii) the commonly made assumption of full-correlation of the interference, which greatly reduces modeling complexity, was shown to be considerably pessimistic compared to the exact result (up to roughly $30\%$ higher outage probability, depending on the path loss exponent); (iii) neglecting the spatial correlation significantly overestimates the true performance; (iv) the outage probability slope is not increased by adding multiple antennas which is due to interference correlation effects.

The CDF of $\mathtt{SIR}$ for the case of more than two antennas was also characterized using bounds. These bounds were then applied to characterize the critical density of simultaneous transmissions given an outage probability constraint as a function of the number of antennas. We concluded the analysis by showing a first-order approximation of the true critical density scaling, indicating a square-root dependence on the number of antennas.

While the proposed bounds are fairly simple, they cannot recover the true $\mathtt{SIR}$-CDF scaling for large number of receive antennas. An extension toward characterizing the $\mathtt{SIR}$ of MRC for an arbitrary number of antennas is hence a promising future direction. Analyzing the performance of MRC under different channel fading and interference geometry assumptions could also be an area of future research. 

\appendix
\subsection{Proof of Theorem~\ref{thm:cdf_general_two_antennas}}\label{ap:general_two_antennas}
Conditioning on $\Phi$ as well as on the fading gains of the second summand ($\mathsf{g}_{2}$ and $\mathbf{h}_{2}$), we can rewrite $\mathbb{P}(\mathtt{SIR}\geq T)$ in \eqref{eq:sir_ccdf} as
\begin{IEEEeqnarray}{rCl}
\mathbb{E}_{\Phi,\mathsf{Z}}\left[\mathbb{P}\left( \frac{\mathsf{g}_{1}d^{-\alpha}}{\sum\limits_{\mathsf{x}_{i}\in\Phi}\mathsf{h}_{1,i}\|\mathsf{x}_{i}\|^{-\alpha}}\geq T-\mathsf{Z}\,\Big|\,\Phi,\mathsf{Z}\right)\right],\IEEEeqnarraynumspace\label{eq:cond_phi_z}
\end{IEEEeqnarray}
where we define the auxiliary variable 
\begin{IEEEeqnarray}{rCl}
\mathsf{Z}=\frac{\mathsf{g}_{2}d^{-\alpha}}{\sum\limits_{\mathsf{x}_{i}\in\Phi}\mathsf{h}_{2,i}\|\mathsf{x}_{i}\|^{-\alpha}}.
\end{IEEEeqnarray}
Note that we recover the single-antenna result ($N=1$) be setting $\mathsf{Z}\equiv0$. Since the fading gains are exponentially distributed, the conditional probability in \eqref{eq:cond_phi_z} can be computed as
\begin{IEEEeqnarray}{rCl}
&&\hspace{-1cm}\mathbb{P}\left( \frac{\mathsf{g}_{1}d^{-\alpha}}{\sum\limits_{\mathsf{x}_{i}\in\Phi}\mathsf{h}_{1,i}\|\mathsf{x}_{i}\|^{-\alpha}}\geq T-\mathsf{Z}\,\big|\,\Phi,\mathsf{Z}\right)\IEEEnonumber\\
&=&\mathbb{P}\left( \mathsf{g}_{1}\geq d^{\alpha}(T-\mathsf{Z})\sum\limits_{\mathsf{x}_{i}\in\Phi}\mathsf{h}_{1,i}\|\mathsf{x}_{i}\|^{-\alpha}\,\big|\,\Phi,\mathsf{Z}\right)\IEEEnonumber\\
&=&\mathbb{E}_{\mathbf{h}_1}\left[ \exp\left(-d^{\alpha}(T-\mathsf{Z})^{+}\sum\limits_{\mathsf{x}_{i}\in\Phi}\mathsf{h}_{1,i}\|\mathsf{x}_{i}\|^{-\alpha}\,\big|\,\Phi,\mathsf{Z}\right)\right]\IEEEnonumber\\
&=& \prod\limits_{\mathsf{x}_{i}\in\Phi}\mathbb{E}_{\mathsf{h}_{1,i}}\left[\exp\left(-\mathsf{h}_{1,i}\|\mathsf{x}_{i}\|^{-\alpha}d^{\alpha}(T-\mathsf{Z})^{+}\right)\,|\,\Phi,\mathsf{Z}\right]\IEEEnonumber\\
&=&\prod\limits_{\mathsf{x}_{i}\in\Phi}\frac{1}{1+\|\mathsf{x}_{i}\|^{-\alpha}d^{\alpha}(T-\mathsf{Z})^{+}},\IEEEeqnarraynumspace\label{eq:inner_prob}
\end{IEEEeqnarray}
where $(z)^{+}=z$ if $z>0$ and zero otherwise. Inserting \eqref{eq:inner_prob} back into \eqref{eq:cond_phi_z}, we obtain
\begin{IEEEeqnarray}{rCl}	\mathbb{E}_{\Phi,\mathsf{Z}}\left[\prod\limits_{\mathsf{x}_{i}\in\Phi}\frac{1}{1+\|\mathsf{x}_{i}\|^{-\alpha}d^{\alpha}(T-\mathsf{Z})^{+}}\right].\IEEEeqnarraynumspace\label{eq:cond_phi_z2}
\end{IEEEeqnarray}
To evaluate the expectation in \eqref{eq:cond_phi_z2}, the PDF of $\mathsf{Z}$ conditional on $\Phi$ is first needed. It can be obtained in a similar way:
\begin{IEEEeqnarray}{rCl}
	\mathbb{P}\left(\mathsf{Z}\geq z\,|\,\Phi\right)&=& \mathbb{P}\left(\frac{\mathsf{g}_{2}d^{-\alpha}}{\sum\limits_{\mathsf{x}_{i}\in\Phi}\mathsf{h}_{2,i}\|\mathsf{x}_{i}\|^{-\alpha}}\geq z\,\Big|\,\Phi\right)\IEEEnonumber\\
	&=&\prod\limits_{\mathsf{x}_{i}\in\Phi}\frac{1}{1+z\|\mathsf{x}_{i}\|^{-\alpha}d^{\alpha}}.\IEEEeqnarraynumspace\label{eq:z_ccdf}
\end{IEEEeqnarray}
Differentiating $1-\prod\limits_{\mathsf{x}_{i}\in\Phi}\frac{1}{1+z\|\mathsf{x}_{i}\|^{-\alpha}d^{\alpha}}$ in \eqref{eq:z_ccdf} with respect to $z$, we obtain the PDF
\begin{IEEEeqnarray}{rCl}
	f_{\mathsf{Z}|\Phi}(z) &=& \sum\limits_{\mathsf{x}_{i}\in\Phi}\frac{\|\mathsf{x}_{i}\|^{-\alpha}d^{\alpha}}{(1+z\|\mathsf{x}_{i}\|^{-\alpha}d^{\alpha})^2}\prod\limits_{\substack{\mathsf{x}_{j}\in\Phi\\\mathsf{x}_{j}\neq\mathsf{x}_{i}}}\frac{1}{1+z\|\mathsf{x}_{j}\|^{-\alpha}d^{\alpha}}\IEEEnonumber\\
&=&\sum\limits_{\mathsf{x}_{i}\in\Phi}\frac{\|\mathsf{x}_{i}\|^{-\alpha}d^{\alpha}}{1+z\|\mathsf{x}_{i}\|^{-\alpha}d^{\alpha}}\,\prod\limits_{\mathsf{x}_{j}\in\Phi}\frac{1}{1+z\|\mathsf{x}_{j}\|^{-\alpha}d^{\alpha}},\IEEEeqnarraynumspace
\end{IEEEeqnarray}
where the second equality follows from the fact $a^2(b\cdot c)+b^2(a\cdot c)+c^2(a\cdot b)=(a+b+c)\cdot(a\cdot b\cdot c)$. Hence, we can rewrite \eqref{eq:cond_phi_z} as
\begin{IEEEeqnarray}{rCl}
&&\hspace{-.5cm}\mathbb{P}(\mathtt{SIR}\geq T)\IEEEnonumber\\
&=& \int_{0}^{\infty}\mathbb{E}_{\Phi}\left[\prod\limits_{\mathsf{x}_{j}\in\Phi}\frac{1}{1+\|\mathsf{x}_{j}\|^{-\alpha}d^{\alpha}(T-z)^{+}}\,f_{\mathsf{Z}|\Phi}(z)\right]\,\mathrm dz\IEEEnonumber\\
&=&\int_{0}^{\infty}\mathbb{E}_{\Phi}\left[\sum\limits_{\mathsf{x}_{i}\in\Phi}\frac{\|\mathsf{x}_{i}\|^{-\alpha}d^{\alpha}}{1+z\|\mathsf{x}_{i}\|^{-\alpha}d^{\alpha}}\right.\IEEEnonumber\\
&&\hspace{.4cm}\left.\times\prod\limits_{\mathsf{x}_{j}\in\Phi}\frac{1}{1+\|\mathsf{x}_{j}\|^{-\alpha}d^{\alpha}(T-z)^{+}}\,\frac{1}{1+z\|\mathsf{x}_{j}\|^{-\alpha}d^{\alpha}}\right]\,\mathrm dz\IEEEnonumber\\
&=&\int_{0}^{\infty}\hspace{-.2cm}\mathbb{E}_{\Phi}\left[\sum\limits_{\mathsf{x}_{i}\in\Phi}\frac{\|\mathsf{x}_{i}\|^{-\alpha}d^{\alpha}}{(1+z\|\mathsf{x}_{i}\|^{-\alpha}d^{\alpha})^2}\,
\frac{1}{1+\|\mathsf{x}_{i}\|^{-\alpha}d^{\alpha}(T-z)^{+}}\right.\IEEEnonumber\\
&&\times\left.\prod\limits_{\substack{\mathsf{x}_{j}\in\Phi\\\mathsf{x}_{j}\neq\mathsf{x}_{i}}}\frac{1}{1+\|\mathsf{x}_{j}\|^{-\alpha}d^{\alpha}(T-z)^{+}}\,\frac{1}{1+z\|\mathsf{x}_{j}\|^{-\alpha}d^{\alpha}}\right]\hspace{-.05cm}\mathrm dz.\IEEEeqnarraynumspace
\end{IEEEeqnarray}
By\cite[Theorem~8.9]{HaenggiBook}, we have for any measurable function $g(x)$
\begin{IEEEeqnarray}{rCl}
	\mathbb{E}\left[\sum\limits_{\mathsf{x}_{i}\in\Phi}g(\mathsf{x}_{i},\Phi\setminus\{\mathsf{x}_{i}\})\right]=\int_{\mathbb{R}^2}\mathbb{E}^{!x}\left[g(x,\Phi)\right]\,\lambda\,\mathrm dx,\IEEEeqnarraynumspace
\end{IEEEeqnarray}
where $\mathbb{E}^{!x}$ is the expectation with respect to the reduced Palm probability $\mathbb{P}^{!x}$. By Slivnyak's Theorem\cite[Theorem~8.10]{HaenggiBook}, conditioning a PPP on having a point at $x$ does not change the statistical law of the rest of the process, i.e., $\mathbb{P}^{!x}\equiv\mathbb{P}$, and hence $\mathbb{E}^{!x}\equiv\mathbb{E}$. Thus,
\begin{IEEEeqnarray}{rCl}
&&\hspace{-.5cm}\mathbb{P}(\mathtt{SIR}\geq T)\IEEEnonumber\\
	&=& \int_{0}^{\infty}\int_{\mathbb{R}^2}\frac{\lambda\|x\|^{-\alpha}d^{\alpha}}{(1+z\|x\|^{-\alpha}d^{\alpha})^2}\,\frac{1}{1+\|x\|^{-\alpha}d^{\alpha}(T-z)^{+}}\IEEEnonumber\\
&&\hspace{-.3cm}\times\mathbb{E}\bigg[\prod\limits_{\mathsf{x}_{j}\in\Phi}\frac{1}{1+z\|\mathsf{x}_{j}\|^{-\alpha}d^{\alpha}}\,\frac{1}{1+\|\mathsf{x}_{j}\|^{-\alpha}d^{\alpha}(T-z)^{+}}\bigg]\mathrm dx\mathrm dz,\IEEEnonumber\\
\end{IEEEeqnarray}
where the expectation can be computed using the probability generating functional for stationary PPPs $\mathbb{E}\left[\prod_{i}v(\mathsf{x}_{i})\right]=\exp(-\lambda\int_{\mathbb{R}^2}(1-v(x))\,\mathrm dx)$ for any non-negative function $v(x)$\cite{stoyan95}. This concludes the proof.\qed

\subsection{Proof of Proposition~\ref{prop:minmax_bounds}}\label{ap:prop_minmax_bounds}
By construction of the $\mathsf{h}_{\min,i}$, the inequality on the left-hand side of \eqref{eq:minmax_bounds} follows from the fact that $\sum_{\mathsf{x}_{i}\in\Phi}\mathsf{h}_{n,i}\|\mathsf{x}_{i}\|^{-\alpha}\geq\sum_{\mathsf{x}_{i}\in\Phi}\mathsf{h}_{\min,i}\|\mathsf{x}_{i}\|^{-\alpha}$ for all $n\in[1,\ldots,N]$. The inequality on the right-hand side of \eqref{eq:minmax_bounds} is in the inverse direction since the construction of the $\mathsf{h}_{\max,i}$ implies $\sum_{\mathsf{x}_{i}\in\Phi}\mathsf{h}_{n,i}\|\mathsf{x}_{i}\|^{-\alpha}\geq\sum_{\mathsf{x}_{i}\in\Phi}\mathsf{h}_{\max,i}\|\mathsf{x}_{i}\|^{-\alpha}$ for all $n\in[1,\ldots,N]$. Using Lemma~\ref{lem:gamma_interference}, the two expressions $\mathbb{P}\left(\mathtt{SIR}_{\max}\leq T\right)$ and $\mathbb{P}\left(\mathtt{SIR}_{\min}\leq T\right)$ can be written in terms of the derivatives of the Laplace transform of the interference, which is computed next. For the $\max$-case, we can directly apply the definition of the Laplace transform of Poisson shot noise\cite{baccelli06} and evaluate the integral over $\mathbb{R}^2$ first which finally yields the result. For the $\min$-case, we first note that $\mathsf{h}_{\min}$ is exponentially distributed with parameter $N$ and then apply the same procedure as for the $\max$-case.\qed

\subsection{Proof of Theorem~\ref{thm:sc-do}}\label{ap:sc-do}
For calculating
	\begin{IEEEeqnarray}{rCl}
		\lim_{\lambda\to 0} \frac{\log \mathbb{P}(\mathtt{SIR}\leq T)}{\log \lambda}&=&\lim_{\lambda\to 0}\frac{\log\left(1-\mathbb{P}(\mathtt{SIR}> T)\right)}{\log \lambda},\IEEEeqnarraynumspace
		\end{IEEEeqnarray}
		it is necessary to characterize $\mathbb{P}(\mathtt{SIR}> T)$ as $\lambda\to0$. 
		Using \eqref{eq:general_two_antennas}, it can be shown that as $\lambda\to0$,
		\begin{IEEEeqnarray}{rcl}
			&&2\pi\lambda\int_{0}^{T}C(z,T)\int_{0}^{\infty}\frac{r^{-\alpha+1}d^{\alpha}}{(1+zr^{-\alpha}d^{\alpha})^2}\frac{\mathrm dr\,\mathrm dz}{1+r^{-\alpha}d^{\alpha}(T-z)}\IEEEnonumber\\
			&&\to \lambda A_{1}+O(\lambda^2),
		\end{IEEEeqnarray}
		where $A_{1}=2\pi\int_{0}^{T}\int_{0}^{\infty}\frac{r^{-\alpha+1}d^{\alpha}}{(1+zr^{-\alpha}d^{\alpha})^2}\,
	\frac{\mathrm dr\,\mathrm dz}{1+r^{-\alpha}d^{\alpha}(T-z)}$, and similarly,
		\begin{IEEEeqnarray}{rCl}
	   &&2\pi\lambda\hspace{-.1cm}\int_{T}^{\infty}\hspace{-.3cm}C(z,T)\hspace{-.1cm}\int_{0}^{\infty}\hspace{-.1cm}\frac{r^{-\alpha+1}d^{\alpha}\,\mathrm dr\,\mathrm dz}{(1+zr^{-\alpha}d^{\alpha})^2}\to 1-\lambda A_{2}+O(\lambda^2),\IEEEeqnarraynumspace
	   \end{IEEEeqnarray}
		where $A_{2}=\frac{2}{\alpha}\pi^2d^2T^{\frac{2}{\alpha}}\text{csc}\left(\frac{2\pi}{\alpha}\right)$. The first part can be verified by the dominated convergence theorem while the second part follows from directly evaluating all three integrals. Hence, $\log\left(1-\mathbb{P}(\mathtt{SIR}> T)\right)\to\log(\lambda(A_{2}-A_{1})+O(\lambda^2))$ as $\lambda\to0$. The desired scaling is obtained only if the linear term inside the $\log$-function is non-vanishing, i.e., $A_{2}-A_{1}>0$. This can be checked as follows
		\begin{IEEEeqnarray}{rCl}
			&&\hspace{-.5cm}(A_{2}-A_{1})\frac{\alpha}{2\pi d^2}T^{-\frac{2}{\alpha}}\IEEEnonumber\\
			&\overset{\text{(a)}}{=}&\pi \text{csc}\left(\frac{2\pi}{\alpha}\right)-\frac{1}{T}\int_{0}^{T}\int_{0}^{\infty}\frac{t^{-\frac{2}{\alpha}}}{(1+tz/T)^2}\,\frac{\mathrm dt\,\mathrm dz}{1+t(1-z/T)}\IEEEnonumber\\
			&\overset{\text{(b)}}{=}&\pi \text{csc}\left(\frac{2\pi}{\alpha}\right)-\int_{0}^{1}\int_{0}^{\infty}\frac{t^{-\frac{2}{\alpha}}}{(1+ts)^2}\,\frac{\mathrm dt\,\mathrm ds}{1+t(1-s)}\IEEEnonumber\\
			&\overset{\text{(c)}}{=}&\pi \text{csc}\left(\frac{2\pi}{\alpha}\right)-\int_{0}^{\infty}\int_{0}^{1}\frac{t^{-\frac{2}{\alpha}}}{(1+ts)^2}\,\frac{\mathrm ds\,\mathrm dt}{1+t(1-s)}\IEEEnonumber\\
			&>&\pi \text{csc}\left(\frac{2\pi}{\alpha}\right)-\int_{0}^{\infty}\int_{0}^{1}\frac{t^{-\frac{2}{\alpha}}}{(1+ts)^2}\,\mathrm ds\,\mathrm dt\IEEEnonumber\\
			&=&\pi \text{csc}\left(\frac{2\pi}{\alpha}\right)-\underbrace{\int_{0}^{\infty}\frac{t^{-\frac{2}{\alpha}}}{1+t}\,\mathrm dt}_{\pi \text{csc}(\frac{2\pi}{\alpha})}=0,
	\end{IEEEeqnarray}
	where (a) follows from the substitution $T(d/r)^{\alpha}\to t$, (b) follows from the substitution $z/T\to s$ and (c) is obtained by swapping the order of integration. Therefore, the scaling is $\log\left(1-\mathbb{P}(\mathtt{SIR}> T)\right)\to\log\lambda+\log(A_{2}-A_{1})$, and hence the SC-DO is $\Delta=\lim\limits_{\lambda\to0}\frac{\log\lambda}{\log\lambda}+\frac{\log(A_{2}-A_{1})}{\log\lambda}=1$.\qed

\IEEEtriggeratref{18}

\end{document}